\documentstyle [12pt] {article}

\newcommand{\beq}{\begin{equation}}
\newcommand{\eeq}{\end{equation}}
\newcommand{\beqs}{\begin{eqnarray}}
\newcommand{\eeqs}{\end{eqnarray}}
\newcommand{\prl}{Phys. Rev. Lett.}
\newcommand{\prd}{Phys. Rev. D}
\newcommand{\npb}{Nucl. Phys. B}
\newcommand{\plb}{Phys. Lett. B}

\begin{document}

\baselineskip 7.5mm

\begin{flushright}
\begin{tabular}{l}
ITP-SB-93-58    \\
September, 1993
\end{tabular}
\end{flushright}

\vspace{8mm}
\begin{center}
{\Large \bf  General Determination of Phases in }\\
\vspace{4 mm}
{\Large \bf Quark Mass Matrices}

\vspace{4mm}
\vspace{16mm}

Alexander Kusenko\footnote{email: sasha@max.physics.sunysb.edu}
and Robert Shrock\footnote{email: shrock@max.physics.sunysb.edu}

\vspace{6mm}
Institute for Theoretical Physics  \\
State University of New York       \\
Stony Brook, N. Y. 11794-3840  \\

\vspace{20mm}

{\bf Abstract}
\end{center}

We construct new invariants and give several theorems which determine in
general
(i) the number of physically meaningful phases in quark mass matrices and
(ii) which elements of these matrices can be rendered real by rephasings.
We illustrate our results with simple models.

\vspace{35mm}

\pagestyle{empty}
\newpage

\pagestyle{plain}
\pagenumbering{arabic}

   Understanding fermion masses and quark mixing remains one of
the most important outstanding problems in particle physics.  In an effort to
gain insight into this problem, many studies of simple models of quark mass
matrices have been carried out over the years.  The phases in these mass
matrices play an essential role in the Kobayashi-Maskawa (KM)
mechanism~\cite{km} for CP violation \cite{add}.
A given model is characterized by
the number of parameters (amplitudes and phases) which specify the quark mass
matrices. Thus, a very important problem is to determine, for any model,
how many physically meaningful complex phases (i.e. phases $\ne 0$ or $\pi$)
occur in the quark mass matrices and which elements of these matrices
can be made real by
rephasings of quark fields.  Surprisingly, there is no general solution to
this question in the literature.  In this Letter we shall present a
general solution and apply our results to several models.

     The quark mass terms can be written in
terms of the SU(3) $\times$ SU(2) $\times$ U(1) fields as
\beq
-{\cal L}_m = \sum_{j,k=1}^3 \Bigl [ v_u \bar Q_{1 j L} Y_{jk}^{(u)} u_{k R}
+ v_d \bar Q_{2 j L} Y_{jk}^{(d)} d_{k R} \Bigr ] + h.c.
\label{massterm}
\eeq
where $j,k$ denote generation indices, $Q_{1 L} = (^u_d)_L$,
$Q_{2 L} = (^c_s)_L$, $Q_{3 L} = (^t_b)_L$; the first
subscript on $Q_{a,j L}$ is the SU(2) index; $u_{1 R} = u_R$, $u_{2 R} = c_R$,
and so forth for $d_{k R}$ \cite{leptons}.  Here we assume three generations of
standard model fermions.  $Y^{(u)}$ and $Y^{(d)}$ are
the Yukawa matrices in the up and down quark sectors whose diagonalization
yields the mass eigenstates $u_{jm}$ and $d_{jm}$.  The $v_u$ and $v_d$ are
mass parameters (Higgs vacuum expectation values in theories with fundamental
Higgs).

    To count the number of physically meaningful phases, we rephase the
fermion fields so as to remove all possible phases in the $Y^{(f)}$.  We
consider the quark sector first.  Here, one can perform the rephasings defined
by
\beq
Q_{j L} = e^{- i \alpha_j} Q_{j L}'
\label{qrephase}
\eeq
\beq
u_{j R} = e^{i \beta^{(u)}_j} u_{j R}'
\label{urephase}
\eeq
\beq
d_{j R} = e^{i \beta^{(d)}_j} d_{j R}'
\label{drephase}
\eeq
for $j=1,2,3$.  In terms of the primed (rephased) fermion fields, the Yukawa
matrices have elements
\beq
Y_{jk}^{(f) \prime} = e^{i(\alpha_j + \beta^{(f)}_k)}Y_{jk}^{(f)}
\label{yfrephased}
\eeq
for $f=u,d$.
Thus, if $Y^{(f)}$ has $N_f$ nonzero, and, in general, complex elements, then
the $N_f$ equations for making these elements real are
\beq
\alpha_j + \beta^{(f)}_k = -arg(Y_{jk}^{(f)}) + \eta^{(f)}_{jk}\pi
\label{yfrephaseq}
\eeq
for $f=u,d$, where the set $\{jk\}$ runs over each of these nonzero elements,
and $\eta^{(f)}_{jk} = 0$ or $1$ \cite{neg}.
Let us define the 9-dimensional vector of fermion field phases
\beq
v = (\{\alpha_i \}, \{ \beta^{(u)}_i \}, \{\beta^{(d)}_i\})^T
\label{pvector}
\eeq
where $\{\alpha_i\} \equiv \{\alpha_1, \alpha_2, \alpha_3\}$, etc.,
and
\beq
w=(\{arg(Y^{(u)})+\eta^{(u)}_{jk}\pi \},\{\{arg(Y^{(d)})+\eta^{(d)}_{jk} \})^T
\label{qvector}
\eeq
of dimension equal to the number of rephasing equations $N_{eq} = N_u + N_d$.
We can then write (\ref{yfrephaseq}) for $f=u,d$ as
\beq
T v = w
\label{aeq}
\eeq
which defines the $N_{eq}$-row by 9-column matrix $T$.

    We first note that $rank(T) \le 8$.  This is proved by ruling out the
only other possibility, i.e. $rank(T)=9$.  The reason that $rank(T)$ cannot
have its apparently maximal value is that one overall rephasing has no effect
on the Yukawa interaction, namely the U(1) generated by
(\ref{qrephase})-(\ref{drephase}) with $-\alpha_i =
\beta^{(u)}_j = \beta^{(d)}_k$ for all $i,j,k$.

   Our first main theorem is:  The number of unremovable
phases $N_p$ in $Y^{(u)}$ and $Y^{(d)}$ is
\beq
N_p = N_{eq}- rank(T)
\label{np}
\eeq
This is proved as follows.  Let $rank(T)=r_T$.  Then one can delete
$N_{eq}-r_T$ rows from the matrix $A$, i.e. not attempt to remove the phases
from the corresponding elements of the $Y^{(f)}$, $f=u,d$.  For the remaining
$r_T$ equations, one moves a subset of $9-(N_{eq}-r_T)$ phases in $v$ to the
right-hand side of (\ref{yfrephaseq}),
thus including them in a redefined $\bar w$.
This yields a set of $r_T$ linear equations in $r_T$ unknown phases, denoted
$\bar v$. We write this as $\bar T \bar v = \bar w$.  Since
by construction $rank(\bar T)=r_T$, $\bar T$ is invertible, so that one can
now solve for the $r_T$ fermion rephasings in $\bar v$ which render $r_T$ of
the $N_{eq}$ complex elements real. Hence there are $N_{eq}-r_T$ remaining
phases in the $Y^{(f)}$, as claimed. $\Box$.

    Some comments are in order.  First, as is clear from our proof, the result
(\ref{np}) does not depend on whether or not $Y^{(f)}_{jk} = Y^{(f)}_{kj}$,
and hence making $Y_{f}$ (complex) symmetric does not, in
general, result in any reduction in $N_p$.  Second, if one of
the unremovable phases is put in a given off-diagonal $Y^{(f)}_{pq}$, one may
wish to modify the $qp'th$ equation to read
\beq
\alpha_q + \beta^{(f)}_p = -arg(Y_{qp}^{(f)})-arg(Y_{pq}^{(f)})
\label{yfhermitian}
\eeq
For example, in a model where $|Y^{(f)}_{pq}|=|Y^{(f)}_{qp}|$, this would
yield $Y_{pq}^{(f)*} = Y_{qp}^{(f)}$ for this pair $pq$.  The modification in
(\ref{yfhermitian}) has no effect on the counting of phases.

   A fundamental question concerns which elements of $Y^{(u)}$ and $Y^{(d)}$
can be made real by fermion rephasings.  This is connected with the issue of
which rows are to be removed from $T$ to obtain $\bar T$, i.e.
which nonzero elements of the $Y^{(u)}$ and
$Y^{(d)}$ are left complex.  We present two more theorems which answer
this question.  The general method is to construct all independent complex
rephasing-invariant (wrt. (\ref{qrephase})-(\ref{drephase})) products of
elements of the $Y^{(f)}$, $f=u,d$.  These must involve an even number of such
elements.  Since, by construction, these have arguments $\ne 0,\pi$, each one
implies a constraint which is that the set of $2n$ elements which comprise it
cannot be made simultaneously real by any fermion rephasings.  We thus
construct a set of invariants depending on the up and down quark sectors
individually:
\beq
P^{(f)}_{2n;j_1 k_1,...j_n k_n;\sigma_L} =
\prod_{a=1}^{n}Y^{(f)}_{j_a k_a}Y^{(f)*}_{\sigma_L(j_a) k_a}
\label{pgeneral}
\eeq
where $f=u,d$, and $\sigma_L$ is an element of the permutation group $S_n$.
Secondly, we construct a set of invariants connecting the up and down quark
sectors:
\beq
Q^{(s,t)}_{2n;\{j\},\{k\},\{m\};\sigma_L,\sigma_u;\sigma_d} =
(\prod_{a=1}^{s}Y^{(u)}_{j_a k_a})
(\prod_{b=1}^{t}Y^{(d)}_{j_{s+b}m_b})
(\prod_{c=1}^{s}Y^{(u)*}_{\sigma_L(j_c)\sigma_u(k_c)})
(\prod_{e=1}^{t}Y^{(d)*}_{\sigma_L(j_{s+e})\sigma_d(m_e)})
\label{qgeneral}
\eeq
where $s, t \ge 1$, $s+t=n$, $\sigma_L \in S_n$, $\sigma_u \in S_s$, and
$\sigma_t \in S_t$.
At quartic order, $2n=4$, if $\sigma$ in (\ref{pgeneral}) equals the
transposition $\tau$, we obtain the complex invariants
\beq
P^{(f)}_{4;j_1 k_1, j_2 k_2; \tau} \equiv P^{(f)}_{j_1 k_1, j_2 k_2} =
Y^{(f)}_{j_1 k_1}Y^{(f)}_{j_2 k_2}Y^{(f)*}_{j_2 k_1}Y^{(f)*}_{j_1 k_2}
\label{p}
\eeq
for $f=u,d$.  At this order, there is only one $Q$-type complex invariant;
this has $s=t=1$, $\sigma_L=\tau$, and we denote it simply as
\beq
Q_{j_1 k_1,j_2 m_1} =
Y^{(u)}_{j_1 k_1}Y^{(d)}_{j_2 m_1}Y^{(u)*}_{j_2 k_1}Y^{(d)*}_{j_1 m_1}
\label{q}
\eeq
Note that
$P^{(f)}_{j_1 k_1, j_2 k_2}=P^{(f)}_{j_2 k_2, j_1 k_1}$,
$P^{(f)}_{j_1 k_1 j_2 k_2}=P^{(f)*}_{j_1 k_2, j_2 k_1}$,
and $Q_{j_1 k_1, j_2 m_1}=Q_{j_2 k_1, j_1 m_1}^*$.

    At order $2n=6$, we find one independent $P$-type complex invariant for
each $f=u,d$, and two independent $Q$-type complex invariants, which we denote
in a simple notation as
\beq
P^{(f)}_{j_1 k_1, j_2 k_2, j_3 k_3} =
Y^{(f)}_{j_1 k_1}Y^{(f)}_{j_2 k_2}Y^{(f)}_{j_3 k_3}
Y^{(f)*}_{j_2 k_1}Y^{(f)*}_{j_3 k_2}Y^{(f)*}_{j_1 k_3}
\label{p6}
\eeq
\beq
Q^{(fff')}_{j_1 k_1, j_2 k_2, j_3 m_1} =
Y^{(f)}_{j_1 k_1}Y^{(f)}_{j_2 k_2}Y^{(f')}_{j_3 m_1}
Y^{(f)*}_{j_1 k_2}Y^{(f)*}_{j_3 k_1}Y^{(f')*}_{j_2 m_1}
\label{q6}
\eeq
where $(fff')=(uud)$ or $(ddu)$.  We find that there are no new constraints
from any invariant of order $\ge 8$.  Note that
$P^{(f)}_{6;j_1 k_1,j_2 k_2, j_3 k_3} =
P^{(f)*}_{j_3 k_2, j_2 k_1, j_1 k_3}$,
$Q^{(fff')}_{6;j_1 k_1, j_2 k_2, j_3 m_1} =
Q^{(fff')*}_{6;j_1 k_2,j_3 k_1, j_2 m_1}$.

   Our theorems are as follows:
For a given model, construct the maximal set of independent complex
invariants of lowest order, whose arguments (phases) are linearly independent.
Denote the number of these by $N_{ia}$. Then
(a) each of these invariants implies a
constraint that the elements contained within it cannot all be made
simultaneously real; (b) this constitutes the complete set of
constraints on which elements of $Y^{(u)}$ and $Y^{(d)}$ can be made
simultaneously real; and hence (c) $N_p = N_{ia}$. Further, (d) there are no
new constraints from any invariant of order $\ge 8$.  Details of proofs are
given in \cite{kslong}.  Usually, the lowest-order nonvanishing complex
invariants are quartic (an exception is given below).

   Note that a set of $N_{inv}$ independent complex invariants of lowest
order, say,
will have arguments which are not in general linearly independent, so $N_{inv}
\ge N_{ia}$.  For each complex invariant of a given order, $X$, $arg(X) =
\sum_{f=u,d}\sum_{j,k}c^{(f)}_{j,k} arg(Y^{(f})_{jk}$, where the sum is
over the $N_{eq}$ complex elements of $Y^{(u)}$ and $Y^{(d)}$. These equations
can be written as $Z \xi = w$, where $\xi$ is the $N_{inv}$-dimensional
vector $\xi = (arg(X_1),...,arg(X_{N_{inv}}))$ and
$Z$ is an $N_{inv}$-row by $N_{eq}$-column matrix.  Then $rank(Z)=N_{ia}$.

    We illustrate our general theorems with some models \cite{models}.
The first is a generalization of models studied in Refs. \cite{rrr} and
\cite{ks}:
\beq
Y^{(u)} =  \left (\begin{array}{ccc}
                  0 & A_{12} & 0 \\
                  A_{21} & A_{22} & 0 \\
                  0 &    0       &  A_{33} \end{array}   \right  )
\label{yu1}
\eeq
\smallskip
\beq
Y^{(d)} =     \left ( \begin{array}{ccc}
                  0 &  B_{12} &  0 \\
                  B_{21} & B_{22} & B_{23} \\
                  0 & B_{32} & B_{33} \end{array}  \right )
\label{yd1}
\eeq
where each of the elements is, in general, complex.  Calculating the $10 \times
9$ matrix, $T$, we find $rank(T)=8$ so from (\ref{np}), there are $N_p=2$
unremovable phases in the $Y_{f}$.  Correspondingly, there are
$N_{inv}=N_{ia}=2 $ complex invariants (with independent arguments):  First,
$P^{(d)}_{22,33}=B_{22}B_{33}B_{32}^*B_{23}^*$ is nonzero, and since,
in general, $arg(P^{(d)}_{22,33}) \ne 0,\pi$, it follows that (i)
 at least one of these phases must
reside among the set $S_1 = \{B_{22},B_{23},B_{32},B_{33}\}$ \cite{spec}, and
(ii) the $2 \times 2$ submatrix in $Y^{(d)}$ formed by $S_1$
(and hence also $Y^{(d)}$ itself) cannot be made hermitian \cite{dis}.
Second, $Q_{12,22}=A_{12}B_{22}A_{22}^*B_{12}^*$ is nonzero, and in general
 $arg(Q_{12,22}) \ne 0,\pi$. Hence, (iii) if one chooses $Y^{(u)}$ real, then
it is not possible to make $B_{12}$ and $B_{22}$ both real.  Conditions
(i)-(iii) are a complete set of rephasing constraints on the $Y^{(f)}$.
These constraints allow both phases to be put in $Y^{(d)}$ and both to be put
in the set $S_1$. If one chooses to make $B_{22}$ and $B_{12}$ real, then one
cannot make $Y^{(u)}$ real, and must assign one phase to $A_{12}$ or $A_{22}$
and the second to $B_{23}$, $B_{32}$ or $B_{33}$.  In accord with our
theorem, although this model has one complex 6'th order invariant,
$Q^{(ddu)}_{32,23,12}=B_{32}B_{23}A_{12}B_{33}^*B_{12}^*A_{22}^*$, its argument
can be expressed in terms of quartic invariant arguments:
$arg(Q^{(ddu)}_{32,23,12}) = arg(Q_{12,22})-arg(P^{(d)}_{22,33})$.

    A second model is given by
\beq
Y^{(u)} =  \left (\begin{array}{ccc}
                  0 & A_{12} & 0 \\
                  A_{21} & 0 & A_{23} \\
                  0 & A_{32} & A_{33} \end{array}   \right  )
\label{yu2}
\eeq
(Fritzsch form \cite{fritzsch}) and $Y^{(d)}$ as in (\ref{yd1}) \cite{rrr}.
We find that the corresponding $11 \times
9$ matrix $T$ has $rank(T)=8$, so $N_p=3$.  There are $N_{inv}=4$ nonzero
independent complex invariants: $P^{(d)}_{22,33}$ as in model 1, together with
$Q_{12,32}=A_{12}B_{32}A_{32}^*B_{12}^*$,
$Q_{23,32}=A_{23}B_{32}A_{33}^*B_{22}^*$, and
$Q_{23,33}=A_{23}B_{33}A_{33}^*B_{23}^*$.  These have only $N_{ia}=3$
independent arguments,
since $arg(P^{(d)}_{22,33})+arg(Q_{23,32})-arg(Q_{23,33})=0$.
The constraints on rephasing from $P^{(d)}_{22,33}$ are (i) and (ii) as above;
and (iii) none of the sets $\{A_{12},A_{32},B_{12},B_{32}\}$,
$\{A_{23},A_{33},B_{22},B_{32}\}$, and $\{A_{23},A_{33},B_{23},B_{33}\}$ can be
made simultaneously real.  In particular, if $Y^{(u)}$ is made real,
then none of the sets $\{B_{12},B_{32}\}$, $\{B_{32},B_{22}\}$, and
$\{B_{23},B_{33}\}$ can be made simultaneously real.

    A third model is given by
\beq
Y^{(u)} =  \left (\begin{array}{ccc}
                  0 & 0 & A_{13} \\
                  0 & A_{22} & 0 \\
                  A_{31} & 0 & A_{33} \end{array}   \right  )
\label{yu3}
\eeq
with $Y^{(d)}$ as in (\ref{yu1}) (this generalizes models considered
in \cite{giudice} and \cite{rrr}).  Here $T$ is $10 \times 9$ with $rank(T)=8$,
so $N_p=2$. We find $N_{inv}=N_{ia}=2$ complex quartic invariants,
$P^{(d)}_{22,33}$ and $Q_{13,32}$. The constraint from $P^{(d)}_{22,33}$ is
given in (i), (ii) above; the constraint from $Q_{13,32}$ is that the set
$\{A_{13},A_{31},B_{12},B_{32}\}$ cannot be made simultaneously real.

   An example of a model for which the lowest order (nonzero) complex invariant
occurs at 6'th order is defined by $Y^{(u)}$ as given by (\ref{yu2}) and
$Y^{(d)}$ by (\ref{yu1}) with $A_{jk} \to B_{jk}$  \cite{dhrn}.
Here, $T$ is $9 \times 9$ with rank 8, whence $N_p=1$.
The $N_{inv}=N_{ia}=1$ 6'th order invariant is
$Q^{(uud)}_{32,23,12}=A_{32}A_{23}B_{12}A_{12}^*A_{33}^*B_{22}^*$. This
yields the constraint that if
$Y^{(u)}$ is made real, then $\{B_{12},B_{22}\}$ cannot be made real, and if
$Y^{(d)}$ is made real, then $\{A_{12},A_{23},A_{32},A_{33}\}$ cannot all be
made real. Further details and applications will be given elsewhere
\cite{kslong}.

   This research was supported in part by NSF grant PHY-93-09888. We thank L.
Lavoura for a discussion on the quartic $P$ invariant.


\begin{thebibliography}{99}

\bibitem{km}{M. Kobayashi and T. Maskawa, Prog. Theor. Phys. {\bf 49}, 652
(1973).}

\bibitem{add}{We consider here the physically relevant case of three
generations of standard model fermions; the quark mixing matrix then
involves one CP-violating phase, $\delta$.  In general, unremovable phase(s)
in the quark Yukawa matrices affect both the CP-violating and CP-conserving
parameters of the quark mixing matrix. Our
analysis does not assume that the KM mechanism is the only source of CPV.}


\bibitem{leptons}{Our methods can also be applied to the leptonic sector, as
we shall discuss elsewhere.}


\bibitem{neg}{The $\eta_{jk}$ term allows for the possibility of making the
rephased element real and negative; this will not affect the counting of
unremovable phases.}

\bibitem{kslong}{A. Kusenko and R. Shrock, ITP-SB-93-62, to appear.}

\bibitem{models}{Although our theorems are general, we note that much recent
work has been done studying models for fermion masses and quark mixing in
supersymmetric grand unified theories (SGUT's) with evolution equations
given by the minimal supersymmetric standard model (MSSM); see, e.g.,
Refs. \cite{r}-\cite{ks} and references to earlier work therein.  These
typically start with $|Y^{(f)}_{jk}|=|Y^{(f)}_{kj}|$, which can be achieved
naturally in SO(10) and minimizes parameters.
Discrete symmetries which could produce zeroes can arise from an
underlying string theory, but string theories
do not generically yield (simple) grand unified groups
or symmetric Yukawa matrices.}


\bibitem{r}{H. Arason, D. Casta\~no, B. Keszthelyi, S. Mikaelian, E. Piard, P.
Ramond, and B. Wright \prl {\bf 67}, 2933 (1991); \prd {\bf 46}, 3945
(1992); H. Arason, D. Casta\~no, P. Ramond, and E. Piard, \prd {\bf 47}, 232
(1993).}

\bibitem{dhr}{S. Dimopoulos, L. Hall, and S. Raby, \prd  {\bf 45}, 4195
(1992); G. Anderson, S. Raby, S. Dimopoulos, L. Hall and G. Starkman,
 LBL-33531 (hep-ph/9308333).}

\bibitem{bbo}{V. Barger, M. Berger, and P. Ohmann, \prd {\bf 47}, 1093 (1992).}

\bibitem{rrr}{P. Ramond, R. Roberts, and G. G. Ross,
RAL-93-010/UFIFT-93-010 (hep-ph/9303320) give a comprehensive list of five
viable models
$M_j$, $j=1-5$ of quark Yukawa matrices with $|Y^{(f)}_{jk}|=|Y^{(f)}_{kj}|$,
$f=u,d$.  Denote the generalization to nonsymmetric $Y^{(f)}$ as  $M_j'$;
this does not affect the counting or allowed placement of the phases. Models
$1-3$ in the text are $M_j'$, $j=1-3$.}

\bibitem{ks}{A. Kusenko and R. Shrock, ITP-SB-93-37 (hep-ph/9307344) studied a
SGUT model for quark and lepton Yukawa matrices which gives a new realization
of the Georgi-Jarlskog mass relation $m_d=3m_e$, $m_s=m_\mu/3$, $m_b=m_\tau$;
the quark part is $M_1$ with $|B_{22}|=|B_{23}|$.}

\bibitem{spec}{In the model of Ref. \cite{ks}, the
unremovable phase in the set $S_1$ did not significantly help the fit to
data and hence was taken to be zero to minimize the number of parameters.}

\bibitem{dis}{These results differ with statements in Ref. \cite{rrr} that
the $Y^{(f)}$ can be made hermitian and with the counting of phases given
there.  For the $M_j'$, $j=1-5$, we find the values of $N_p$ are
respectively $2,3,2,2,2$, not $1,2,1,1,1$ as claimed.  Further,
$Y^{(d)}$ cannot in general be made hermitian (nor can the submatrix $S_1$
be made real or hermitian) in $M_j'$, $j=1,2,3$, and $Y^{(u)}$ cannot be
made real (or hermitian) in $M_4'$ or $M_5'$.  We find the
following invariants for the other two models:
$M_4'$: $(P^{(u)}_{22,33},Q_{12,22})$,
$M_5'$: $(P^{(u)}_{22,33},Q_{13,22})$. We have communicated our results to P.
Ramond and G. G. Ross and thank them for discussions.}

\bibitem{fritzsch}{H. Fritzsch, \plb {\bf 73}, 317
(1978); \npb {\bf 155}, 182 (1979).}

\bibitem{giudice}{G. Giudice, Mod. Phys. Lett. A {\bf 7}, 2429 (1992).}


\bibitem{dhrn}{The special case with $|Y^{(f)}_{jk}|=|Y^{(f)}_{kj}|$, $f=u,d$
has has been studied in Refs. \cite{r}-\cite{bbo}.}

\end{thebibliography}
\end{document}